\documentclass[12pt]{article}
\usepackage{eqsection,subeqnarray,indent,amsfonts,latexsym,epsf}

\begin{document}

\bibliographystyle{plain}

\date{April 18, 2000 \\[2mm] 
     to appear in {\em Markov Processes and Related Fields}
     }

\title{\vspace*{-1cm} 
       Dynamic critical behavior of cluster 
       algorithms for 2D Ashkin--Teller and
       Potts models\thanks{Talk presented at the Conference on Inhomogeneous
       Random Systems. Universit\'e de Cergy-Pontoise, January 25, 2000.} 
       }

\author{
  {\small Jes\'us Salas}                                    \\[-0.2cm]
  {\small\it Departamento de F\'{\i}sica Te\'orica}         \\[-0.2cm]
  {\small\it Facultad de Ciencias, Universidad de Zaragoza} \\[-0.2cm]
  {\small\it Zaragoza 50009, SPAIN}                         \\[-0.2cm]
  {\small\tt SALAS@LAUREL.UNIZAR.ES}                        \\[-0.2mm]
  {\protect\makebox[5in]{\quad}}  
  \\
}
\vspace{0.5cm}

\maketitle
\thispagestyle{empty}   

\def\spose#1{\hbox to 0pt{#1\hss}}
\def\ltapprox{\mathrel{\spose{\lower 3pt\hbox{$\mathchar"218$}}
 \raise 2.0pt\hbox{$\mathchar"13C$}}}
\def\gtapprox{\mathrel{\spose{\lower 3pt\hbox{$\mathchar"218$}}
 \raise 2.0pt\hbox{$\mathchar"13E$}}}
\def\inapprox{\mathrel{\spose{\lower 3pt\hbox{$\mathchar"218$}}
 \raise 2.0pt\hbox{$\mathchar"232$}}}


\begin{abstract}
We study the dynamic critical behavior of two algorithms: the Swendsen--Wang 
algorithm for the two-dimensional Potts model with $q=2,3,4$ and a 
Swendsen--Wang--type algorithm for the two-dimensional symmetric Ashkin--Teller
model on the self-dual curve.
We find that the Li--Sokal bound on the autocorrelation time 
$\tau_{{\rm int},{\cal E}} \geq \hbox{const} \times C_H$ is almost, but not
quite sharp. The ratio $\tau_{{\rm int},{\cal E}}/C_H$ appears to tend to 
infinity either as a logarithm or as a small power 
($0.05 \ltapprox p \ltapprox 0.12$). We also show that the exponential 
autocorrelation time $\tau_{{\rm exp},{\cal E}}$ is proportional to the 
integrated autocorrelation time $\tau_{{\rm int},{\cal E}}$.  
\end{abstract}

\bigskip
\noindent
{\bf Running title:} Dynamic Critical Behavior of Cluster Algorithms


\bigskip
\noindent
{\bf AMS classification numbers:} 68W20, 82B20, 82B26, 82B80. 

\clearpage

\newcommand{\var}{\hbox{\rm var}}

\newcommand{\be}{\begin{equation}}
\newcommand{\ee}{\end{equation}}
\newcommand{\<}{\langle}
\renewcommand{\>}{\rangle}
\newcommand{\widebar}{\overline}
\def\reff#1{(\protect\ref{#1})}
\def\spose#1{\hbox to 0pt{#1\hss}}
\def\ltapprox{\mathrel{\spose{\lower 3pt\hbox{$\mathchar"218$}}
 \raise 2.0pt\hbox{$\mathchar"13C$}}}
\def\gtapprox{\mathrel{\spose{\lower 3pt\hbox{$\mathchar"218$}}
 \raise 2.0pt\hbox{$\mathchar"13E$}}}
\def\textprime{${}^\prime$}
\def\proof{\par\medskip\noindent{\sc Proof.\ }}
\def\qed{\hbox{\hskip 6pt\vrule width6pt height7pt depth1pt \hskip1pt}\bigskip}
\def\proofof#1{\bigskip\noindent{\sc Proof of #1.\ }}
\def\half{ {1 \over 2} }
\def\third{ {1 \over 3} }
\def\twothird{ {2 \over 3} }
\def\smfrac#1#2{\textstyle{#1\over #2}}
\def\smhalf{ \smfrac{1}{2} }
\newcommand{\real}{\mathop{\rm Re}\nolimits}
\renewcommand{\Re}{\mathop{\rm Re}\nolimits}
\newcommand{\imag}{\mathop{\rm Im}\nolimits}
\renewcommand{\Im}{\mathop{\rm Im}\nolimits}
\newcommand{\sgn}{\mathop{\rm sgn}\nolimits}
\newcommand{\tr}{\mathop{\rm tr}\nolimits}
\newcommand{\diag}{\mathop{\rm diag}\nolimits}
\newcommand{\Gal}{\mathop{\rm Gal}\nolimits}
\newcommand{\mycup}{\mathop{\cup}}
\def\hboxscript#1{ {\hbox{\scriptsize\em #1}} }
\def\zhat{ {\widehat{Z}} }
\def\phat{ {\widehat{P}} }
\def\qtilde{ {\widetilde{q}} }

\def\scra{\mathcal{A}}
\def\scrb{\mathcal{B}}
\def\scrc{\mathcal{C}}
\def\scrf{\mathcal{F}}
\def\scrg{\mathcal{G}}
\def\scrl{\mathcal{L}}
\def\scro{\mathcal{O}}
\def\scrp{\mathcal{P}}
\def\scrq{\mathcal{Q}}
\def\scrr{\mathcal{R}}
\def\scrs{\mathcal{S}}
\def\scrt{\mathcal{T}}
\def\scrv{\mathcal{V}}
\def\scrz{\mathcal{Z}}

\def\q{{\sf q}}

\def\Z{{\mathbb Z}}
\def\R{{\mathbb R}}
\def\C{{\mathbb C}}
\def\Q{{\mathbb Q}}

\def\T{{\mathsf T}}
\def\H{{\mathsf H}}
\def\V{{\mathsf V}}
\def\D{{\mathsf D}}
\def\J{{\mathsf J}}
\def\P{{\mathsf P}}
\def\QQ{{\mathsf Q}}
\def\RR{{\mathsf R}}

\def\bsigma{\mbox{\protect\boldmath $\sigma$}}
\def\bone{{\mathbf 1}}
\def\vv{{\bf v}}
\def\w{{\bf w}}

\newtheorem{theorem}{Theorem}[section]
\newtheorem{proposition}[theorem]{Proposition}
\newtheorem{lemma}[theorem]{Lemma}
\newtheorem{corollary}[theorem]{Corollary}
\newtheorem{conjecture}[theorem]{Conjecture}


\newenvironment{sarray}{
          \textfont0=\scriptfont0
          \scriptfont0=\scriptscriptfont0
          \textfont1=\scriptfont1
          \scriptfont1=\scriptscriptfont1
          \textfont2=\scriptfont2
          \scriptfont2=\scriptscriptfont2
          \textfont3=\scriptfont3
          \scriptfont3=\scriptscriptfont3
        \renewcommand{\arraystretch}{0.7}
        \begin{array}{l}}{\end{array}}

\newenvironment{scarray}{
          \textfont0=\scriptfont0
          \scriptfont0=\scriptscriptfont0
          \textfont1=\scriptfont1
          \scriptfont1=\scriptscriptfont1
          \textfont2=\scriptfont2
          \scriptfont2=\scriptscriptfont2
          \textfont3=\scriptfont3
          \scriptfont3=\scriptscriptfont3
        \renewcommand{\arraystretch}{0.7}
        \begin{array}{c}}{\end{array}}


\section{Introduction} \label{sec_intro} 

Monte Carlo (MC) simulations \cite{Binder_87,Binder_92,Sokal_Cargese}
have become a standard and powerful tool for gaining new insights into
statistical-mechanical systems and lattice field theories.
However, their practical success is severely limitated by critical 
slowing-down: the autocorrelation time $\tau$, 
which roughly measures the MC time needed to produce a 
``statistically independent'' configuration, diverges near a critical point.   
More precisely, for a finite system of linear size $L$ at criticality, we
expect a behavior $\tau \sim L^z$ for large $L$. The power $z$ is a 
{\em dynamic critical exponent} and it depends on both the system and the
algorithm. 

In single-site MC algorithms (such as single-site Metropolis or 
heat bath), critical slowing-down arises fundamentally from the fact that 
the updates are local; and have a dynamic critical exponent $z\gtapprox 2$. 
This is a severe critical slowing-down: the total 
amount of computer work needed to study a $d$-dimensional lattice of linear 
size $L$ grows $L^2$ times faster than the naive geometrical factor $L^d$. 
To study static and dynamic critical behavior we need high-precision data 
(run lengths $\ltapprox 10^4 \tau$), so in practice, it is very difficult to
obtain high-precision data for large lattices with local algorithms.  
The geometrical factor $L^d$ is unavoidable for the usual MC 
simulations, so the elimination (or the reduction) of the critical slowing down
is the only way to make MC simulations feasible close to a critical point. 

In some cases, a much better dynamic behavior is obtained by allowing
nonlocal moves, such as cluster flips. In particular, the Swendsen--Wang (SW)
algorithm for the $q$-state ferromagnetic Potts model \cite{Swendsen_87}
achieves a significant reduction in $z$ compared to the local algorithms:
numerical experiments suggest that $0.22 \ltapprox z \ltapprox 1$, where
the exact value depends on the dimensionality of the lattice $d$ and on $q$ 
(See Table~\ref{table_potts}).
The most favorable case is the two-dimensional (2D) Ising model ($q=2$), for 
which the numerical data suggest $z = 0.2186\pm0.0068$ 
\cite{Salas_Sokal_Ising} but may also be compatible with a logarithmic 
growth \cite{Heermann_90,Baillie_91,Baillie_92}.  
In other cases, the performance of the SW algorithm is less impressive
(though still quite good): e.g., $z = 0.515 \pm 0.006$ for the 2D 3-state
Potts model \cite{Salas_Sokal_Potts3},
$z = 0.876\pm 0.011$ for the 2D 4-state Potts model
\cite{Salas_Sokal_AT,Salas_Sokal_FSS},
and $z \approx 1$ for the 4D Ising model \cite{Klein_89,Ray_89}.
Clearly, we would like to understand why this algorithm works so well
in some cases and not in others. We hope in this way to obtain new insights
into the dynamics of non-local MC algorithms,
with the ultimate aim of devising new and more efficient algorithms.

%
%
\begin{table}[htb]
\centering
\begin{tabular}{|l|l|l|l|}
\hline\hline 
$d$    &    $q$     & \multicolumn{1}{|c|}{$z_{{\rm int},{\cal E}}$} & 
                      \multicolumn{1}{|c|}{$\alpha/\nu$} \\
\hline\hline 
2      &    2       & $0.2186\pm 0.0068$ \cite{Salas_Sokal_Ising} &  
                      $0\times \log$ \\ 
       &            & $0\times \log$ \cite{Heermann_90,Baillie_91,Baillie_92}& 
                                     \\
\cline{2-4}
       &    3       & $0.515 \pm 0.006$ \cite{Salas_Sokal_Potts3} & 
                      $2/5 = 0.4$ \\
\cline{2-4} 
       &    4       & $0.876 \pm 0.011$ \cite{Salas_Sokal_FSS} & 
                      $1 \times \log^{-3/2} \approx 0.770\pm 0.018$ \\ 
\hline  
3      &    2       & $0.339\pm 0.004$       \cite{Wang_90}     & 
                      $0.1696\pm 0.0046$     \cite{Caselle_97}  \\ 
       &            & $\approx 0.50\pm 0.03$ \cite{Wolff_89,Baillie_92} &  \\
       &            & $0.75\pm 0.01$         \cite{Swendsen_87} &  \\
\hline 
4      &    2       & $\approx 1$           \cite{Klein_89,Ray_89} & 
                      $0 \times \log^{3/2}$ \\
       &            & $0.86\pm 0.02$        \cite{Baillie_92}      & \\
\hline\hline  
\end{tabular}
\caption{\protect\label{table_potts}
Numerical estimates of the dynamical critical exponent
$z_{{\rm int},{\cal E}}$ for Potts models. We also show the values of 
the ratio $\alpha/\nu$. The exact values of $\alpha/\nu$ have been taken from
Ref.~\cite{Baxter,Itzykson_89}. 
}
\end{table}

There is at present no adequate theory for predicting the
dynamic critical behavior of an SW-type algorithm. However, there is one 
rigorous lower bound on $z$ due to Li and Sokal \cite{Li_Sokal}.
The autocorrelation times of the standard (multi-cluster) SW algorithm for 
the ferromagnetic $q$-state Potts model are bounded below by a multiple of 
the specific heat:
\be
  \tau_{{\rm int},{\cal N}}, \;
  \tau_{{\rm int},{\cal E}}, \;
  \tau_{\rm exp}  \;\geq\; {\rm const} \times C_H   \;.
\label{Li_Sokal_bound_CH}
\ee
Here ${\cal N}$ is the bond density [c.f. \reff{def_bond_and_energy}] in the 
SW algorithm, ${\cal E}$ is the energy, and $C_H$ is the specific heat.  
As a result one has
\be
  z_{{\rm int},{\cal N}}, \;
  z_{{\rm int},{\cal E}}, \;
  z_{\rm exp}  \;\geq\; {\alpha \over \nu } \; ,
\label{Li_Sokal_bound}
\ee
where $\alpha$ and $\nu$ are the standard {\em static}\/ critical exponents.
Thus, the SW algorithm for the $q$-state Potts model cannot
{\em completely}\/ eliminate the critical slowing-down in any model
in which the specific heat is divergent at criticality.

The important question is the following: Is the Li--Sokal bound 
\reff{Li_Sokal_bound_CH}/\reff{Li_Sokal_bound} sharp or not? 
An affirmative answer would imply that we could use the bound to predict 
the dynamic critical exponent(s) $z$ given only the static critical exponents 
of the system. There are three possibilities: 

\begin{enumerate} 
\item[i)] 
The bound \reff{Li_Sokal_bound_CH} is {\em sharp}\/ 
(i.e., the ratio $\tau/C_H$ is bounded), so that \reff{Li_Sokal_bound} is an  
{\em equality}\/. 

\item[ii)] 
The bound is {\em sharp modulo a logarithm}\/ 
(i.e., $\tau/C_H \sim \log^p L$ for $p>0$). 

\item[iii)] 
The bound is {\em not sharp}\/ 
(i.e., $\tau/C_H \sim L^p$ for $p>0$), so that \reff{Li_Sokal_bound} is a  
{\em strict inequality}\/.      
\end{enumerate} 

Unfortunately, the empirical situation, even for the simplest cases, is 
far from clear. From Table~\ref{table_potts} we see that the Li--Sokal 
bound is apparently not sharp for the Ising model in $d>2$. However, the
differences $z_{{\rm int},{\cal E}}-\alpha/\nu$ are much smaller in $d=2$. 
A first look at Table~\ref{table_potts} reveals that the Li--Sokal bound 
for the Ising model could only be sharp if the autocorrelation time
grows like a logarithm \cite{Heermann_90,Baillie_91,Baillie_92}. Otherwise,
the bound \reff{Li_Sokal_bound} would be non-sharp. The Li--Sokal bound for
the 3-state Potts model seems to be non-sharp: the difference 
$z_{{\rm int},{\cal E}}- \alpha/\nu = 0.115\pm0.006$ is clearly not 
consistent with zero. Finally, the 4-state Potts model is rather peculiar:
the naive fit to the data, $z_{{\rm int},{\cal E}} = 0.876 \pm 0.011$
\cite{Salas_Sokal_FSS}, is {\em smaller}\/ than the (exactly known)
value of $\alpha/\nu = 1$. The explanation of this paradox is that the 
true leading term in the specific heat has a multiplicative logarithmic 
correction, $C_H \sim L \log^{-3/2} L$
\cite{Nauenberg_Scalapino,Cardy80,Black_Emery,Salas_Sokal_FSS}.
Indeed, a naive power-law fit to the specific heat yields
$\alpha/\nu = 0.770\pm 0.008$ \cite{Salas_Sokal_FSS}, 
consistent with the bound \reff{Li_Sokal_bound}. The Li--Sokal bound would
be sharp modulo a logarithm if $\tau \sim L \log^p L$ with $p \geq -3/2$.
Thus, in $d=2$, $q=2$ and $q=4$ are candidates for a sharp (perhaps modulo a 
logarithm) Li--Sokal bound; but for $q=3$ this bound is apparently non-sharp,  
at least if the numerical estimates are to be trusted. It is important,
however, to proceed cautiously in interpreting these estimates, which are
after all only fits to data from a limited range of $L$ values (typically 
$L\ltapprox 1024$) and which may be significantly biased by corrections to
scaling. In particular, it is extremely difficult to distinguish 
numerically between a logarithm and a small power.\footnote{Actually, 
  if we do not have access to
  many orders of magnitude in $L$, a power-law with a small power $p$ is
  consistent with a logarithm:
  $L^p \approx 1 + p \log L + {\cal O}\left( p^2 \log^2 L \right)$.
}

The above empirical situation motivated in part the study of the 2D 
Ashkin--Teller (AT) model using a SW-type algorithm \cite{Salas_Sokal_AT}.  
This model interpolates between the Ising and the 4-state Potts models. 
The study of the AT model proved to be useful to obtain new insights on the 
sharpness of the Li--Sokal bound. 

In Section~\ref{sec_defs} we will define more rigorously the concept of 
autocorrelation time. In Section~\ref{sec_sw} we will review the SW algorithm 
for the $q$-state ferromagnetic Potts model and the Li--Sokal bound for 
such algorithm. In Section~\ref{sec_AT} we will introduce two different 
SW-type algorithms for the AT model. In Section \ref{sec_results} 
we will analyze the sharpness of the Li--Sokal bound for both cases.  
Finally, in Section~\ref{sec_tau_exp} we will consider the dynamic critical 
behavior of the exponential autocorrelation time.

\section{Autocorrelation times} \label{sec_defs} 

The goal of a MC simulation is to generate random samples of the 
spins $\sigma = \{ \sigma_i \}$ distributed according to a probability 
distribution $\Pi$. This measure is usually Gibbsian  
$\Pi = e^{- {\cal H}(\sigma) } /Z$, 
where ${\cal H}$ is the reduced Hamiltonian of the system and $Z$ is the 
partition function (which is generally unknown). 

To do this, we {\em invent} a transition probability matrix 
$P(\sigma \rightarrow \sigma')$ that tells us the probability of going from a 
spin configuration $\sigma$ to a new configuration $\sigma'$. 
This matrix should be ergodic (i.e., from any spin configuration one can reach 
any other), and should leave $\Pi$ invariant  
$\sum_{\{\sigma\}} \Pi(\sigma) P(\sigma \rightarrow \sigma') = \Pi(\sigma')$. 
The (usually easier to prove) detailed balance 
$\Pi(\sigma)  P(\sigma  \rightarrow \sigma') = 
 \Pi(\sigma') P(\sigma' \rightarrow \sigma )$ 
is a sufficient, but {\em not} a necessary condition for stationarity. 

We then simulate the Markov process defined by the transition matrix $P$, 
starting at some arbitrary initial configuration $\sigma^{(0)}$. 
In this way we generate a random sequence of configurations 
$\sigma^{(1)}, \sigma^{(2)}, \ldots \sigma^{(t)}, \ldots$. 
Given the two properties above (e.g., ergodicity and stationarity of $\Pi$), 
it is guaranteed that this Markov chain converges to the equilibrium 
distribution $\Pi$, irrespective of the initial configuration.  

In every MC simulation two different autocorrelation times can be 
defined. Given an arbitrary observable of the spins $A(\sigma)$, we can 
define its {\em equilibrium} normalized autocorrelation function:
\be
\rho_{AA}(t) = {\left\langle A^{(s)} A^{(s+t)} \right\rangle - 
              \langle A \rangle^2  
                \over 
               \left\langle A^2 \right\rangle - \langle A \rangle^2 
               } 
\label{autocorrelation_function}
\ee 
Typically, one expects that this function decays exponentially 
(e.g., $\rho_{AA}(t) \sim e^{- |t|/\tau}$). This motivates the following 
definition of the exponential autocorrelation time for the observable $A$ 
\be
\tau_{{\rm exp},A} = \lim \sup_{t\rightarrow\infty} 
  { t \over - \log | \rho_{AA}(t) |} 
\label{def_tau_exp_A}
\ee
The autocorrelation time $\tau_{{\rm exp},A}$ measures the 
relaxation time of the slowest mode coupled to the observable $A$. The 
exponential autocorrelation time of the system is the relaxation time of 
the slowest mode of the system 
\be 
\tau_{\rm exp} = \sup_A \tau_{{\rm exp},A}
\label{def_tau_exp}
\ee
This number can be {\em infinite} for some perfectly legal algorithms!
\cite{Sokal_Cargese}.
The integrated autocorrelation time of the observable $A$ is also defined 
in terms of $\rho_{AA}$:
\be
\tau_{{\rm int},A} = {1 \over 2} \sum_{t=-\infty}^\infty \rho_{AA}(t)
\label{def_tau_int_A}
\ee

There are two fundamental (and distinct) issues in dynamical MC simulations: 

\medskip 

\noindent 
1) {\em Initialization bias}. If the initial configuration $\sigma^{(0)}$ is 
not ``characteristic'' of the equilibrium probability distribution $\Pi$, 
the first MC configurations are not distributed according to $\Pi$; 
they will introduce a {\em systematic} error in our MC estimates. 
To eliminate this bias, we have to discard the first $N_0$ iterations such 
that when $t > N_0$ the system has reached the equilibrium distribution $\Pi$.  
But how large should $N_0$ be? One answer is given by the 
exponential autocorrelation time $\tau_{\rm exp}$ \reff{def_tau_exp}. 
This autocorrelation time is a worst-case bound on relaxation to equilibrium.
For most systems, it suffices to discard the first 
$N_0 \gtapprox 20 \tau_{\rm exp}$ iterations. In this case, the deviation from 
equilibrium (in the $l^2$ sense) will be at most 
$e^{-20}$ ($\approx 2\times 10^{-9}$) times the initial deviation from 
equilibrium.  

\medskip

\noindent 
2) {\em Autocorrelation in equilibrium}. Even if the system is in 
equilibrium, the MC configurations are not statistically independent,
but highly correlated. The MC estimate of the mean value 
$\langle A \rangle$ is given by 
\be
  \bar{A} = {1 \over N} \sum_i A(\sigma^{(i)}) \; , 
\ee
where $N$ is the total number of measurements. The variance of such estimator 
can be written as  
\be
\var (\bar{A} ) = {1 \over N^2} \sum_{ij} C_{AA}(i-j) 
                  \approx {C_{AA}(0) \over N} 2 \tau_{{\rm int},A}\; ,  
\label{var_A}
\ee 
when $N \gg \tau_{{\rm int},A}$. Thus, $\var(\bar{A})$ is 
$2 \tau_{{\rm int},A}$ times larger than in independent sampling. 
This means we need a good determination of the integrated autocorrelation 
time \reff{def_tau_int_A} in order to obtain  a reliable estimate of 
$\var(\bar{A})$. In practice, the condition 
$N \gg \tau_{{\rm int},A}$ is replaced by 
$N \gtapprox 10^4 \tau_{{\rm int},A}$ \cite{Madras_Sokal}. 

Close to a second-order phase transition both autocorrelation times 
\reff{def_tau_exp_A}/\reff{def_tau_int_A} diverge as a power law\footnote{
  Close to first-order phase transitions, we expect that the autocorrelation 
  times behave like $\tau \sim e^{ c L^{d-1}}$, where $c$ is a constant. 
  This exponential growth is due to the tunneling among the different pure 
  phases coexisting at the (first-order) phase transition. 
} 
\begin{subeqnarray}
  \tau_{{\rm exp},A} &\sim& \min[L,\xi]^{z_{{\rm exp},A}} \\ 
  \tau_{{\rm int},A} &\sim& \min[L,\xi]^{z_{{\rm int},A}} 
\end{subeqnarray}
The dynamic critical exponents $z_{{\rm exp},A}$ and $z_{{\rm int},A}$ 
are in general different \cite{Sokal_Cargese} (See Section~\ref{sec_tau_exp}).

\section{Swendsen--Wang algorithm for the Potts model} \label{sec_sw}

The SW algorithm \cite{Swendsen_87} is defined for the ferromagnetic 
$q$-state Potts model. This model assigns to each 
lattice site $i$ a spin variable $\sigma_i$ taking values in the set 
$\{1,2,\ldots,q\}$. These spins interact through the reduced Hamiltonian
\be
{\cal H}_{\rm Potts}   \;=\;   - J_{ij} \sum_{\< ij \>}
      (\delta_{\sigma_i,\sigma_j} - 1) \; ,
\label{Potts_Hamiltonian}
\ee
where the sum runs over all the nearest-neighbor pairs $\<ij\>$, and the 
all the coupling constants are positive $J_{ij} \geq 0$ $\forall \< ij \>$.  
The Boltzmann weight of a configuration $\{\sigma\}$ is given by
\be
W_{\rm Potts}(\{\sigma\})   \;=\;
 {1 \over Z} \prod_{\<ij\>} e^{J_{ij} (\delta_{\sigma_i,\sigma_j} - 1)}
   \;=\;   {1 \over Z} \prod_{\<ij\>} 
           ( 1-p_{ij} + p_{ij} \delta_{\sigma_i,\sigma_j} )
\label{Potts_weight}
\ee
where $p_{ij}=1-e^{-J_{ij}}$, and the partition function is given by 
$Z = \sum_{\{\sigma\}} e^{-{\cal H}_{\rm Potts}}$.
The idea behind the SW algorithm \cite{Swendsen_87,Edwards_Sokal,Sokal_Cargese}
is to decompose the Boltzmann weight by introducing new dynamical 
variables $n_{ij}=0,1$ (living on the bonds of the lattice), and to simulate 
the joint model of spin and bond variables by alternately updating one 
set of variables conditional on the other set. The Boltzmann weight of the 
joint model is
\be
W_{\rm joint}(\{\sigma\};\{n\})   \;=\;   {1 \over Z} \prod_{\<ij\>} \left[
   (1-p_{ij}) \delta_{n_{ij},0} + p_{ij} \delta_{\sigma_i,\sigma_j}
         \delta_{n_{ij},1} \right]
   \;.
\label{Potts_joint_weight}
\ee
The marginal distribution of \reff{Potts_joint_weight} with respect to the 
spin variables reproduces the Potts-model Boltzmann weight \reff{Potts_weight}.
The marginal distribution of \reff{Potts_joint_weight} with respect to the 
bond variables is the Fortuin--Kasteleyn 
\cite{Kasteleyn_69,Fortuin_Kasteleyn_72} random-cluster model with parameter 
$q$:
\be
W_{\rm RC}(\{n\})   \;=\;
   {1 \over Z} \left[ \prod_{\<ij\> \colon\; n_{ij}=1} p_{ij} \right]
 \left[ \prod_{\<ij\> \colon\; n_{ij}=0} (1-p_{ij}) \right] q^{{\cal C}(\{n\})}
 \;,
\label{RC_weight}
\ee
where ${\cal C}(\{n\})$ is the number of connected components (including
one-site components) in the graph whose edges are the bonds with
$n_{ij}=1$.

We can also consider the conditional probabilities of the joint
distribution \reff{Potts_joint_weight}.
The conditional distribution $P_{\rm bond} = E(\cdot | \{\sigma\})$ 
of the $\{n\}$ given the $\{\sigma\}$ is as follows: 
independently for each bond $\<ij\>$, one sets
$n_{ij}=0$ when $\sigma_i \neq \sigma_j$, and sets
$n_{ij}=0$ and 1 with probabilities $1-p_{ij}$ and $p_{ij}$ 
when $\sigma_i=\sigma_j$.
The conditional distribution $P_{\rm spin} = E(\cdot | \{n\})$ of 
the $\{\sigma\}$ given the $\{n\}$ is as follows: independently for each 
connected cluster, one sets all the spins $\sigma_i$ in that cluster equal 
to the same value, chosen with uniform probability from the set 
$\{1,2,\ldots,q\}$.

The SW algorithm simulates the joint probability distribution 
\reff{Potts_joint_weight} by alternately applying the two conditional 
distributions just described (i.e., $P_{SW} = P_{\rm bond} P_{\rm spin}$).  

\medskip

\noindent
{\bf Step 1.} Generate a new bond configuration $\{n\}$ from $P_{\rm bond}$. 

\medskip

\noindent
{\bf Step 2.} Generate a new spin configuration $\{\sigma\}$ from 
$P_{\rm spin}$.

\bigskip

The performance of this algorithm has been discussed in Section~\ref{sec_intro} 
(See also Table~\ref{table_potts}). 
Let us now briefly review the proof of the Li--Sokal bound 
\reff{Li_Sokal_bound_CH}/\reff{Li_Sokal_bound} for the homogeneous Potts model
$p_{ij}=p$ \cite{Li_Sokal}. The strategy of the proof is simple: first, we 
choose two ``slow'' observables: the bond and energy densities  
\be
{\cal N} = \sum_{\<ij\>} n_{ij} \; , \qquad 
{\cal E} = \sum_{\<ij\>} \delta_{\sigma_i\sigma_j} \; , 
\label{def_bond_and_energy}  
\ee 
and compute their autocorrelation functions at time lags 0 and 1. 
Then, using some general properties of reversible Markov chains, we will 
deduce lower bounds for the autocorrelation times $\tau_{{\rm int},A}$ 
(for $A = {\cal N}, {\cal E}$) and $\tau_{\rm exp}$.
These will in turn imply lower bounds on the
dynamic critical exponents $z_{{\rm int},A}$ and $z_{\rm exp}$.

{}From \reff{Potts_joint_weight} we can compute the expectation values of 
bond variables $n_{ij}$ conditional to the spin variables. In particular, we
can compute the autocorrelation function of the bond occupation 
${\cal N}$ at time lag 1 \cite{Li_Sokal,Salas_Sokal_Potts3}
\be
\rho_{\cal NN}(1) = 1 - { (1-p) E \over p C_H + (1-p) E } 
                  \geq 1 - {A \over C_H} \; , 
\label{inequality_1}
\ee
where the energy density $E$ and the specific heat $C_H$ are defined by: 
\be
E   = {1 \over V} \< {\cal E} \> \; , \qquad  
C_H = {1 \over V} \var ({\cal E}) \; .  
\ee 
At criticality, $p \rightarrow p_c > 0$ and $E \rightarrow E_c > 0$, so the
constant $A$ does not vanish. 

The correlation functions of ${\cal N}$ under 
$P_{SW} = P_{\rm bond}P_{\rm spin}$ are the same as under the 
positive-semidefinite self-adjoint operator 
$P_{SW}' = P_{\rm spin}P_{\rm bond}P_{\rm spin}$. This implies \cite{Li_Sokal} 
that we have a spectral representation 
\be
\rho_{\cal NN}(t) = \int_0^1 \lambda^{|t|} d\nu(\lambda) \;,  
\ee
with a positive measure $d\nu$. It follows that 
\be 
\rho_{\cal NN}(t) \geq \rho(1)^{|t|} \; . 
\label{inequality_2}
\ee
If we introduce the inequalities \reff{inequality_1}/\reff{inequality_2} 
in the definition of the autocorrelation times 
\reff{def_tau_exp_A}/\reff{def_tau_int_A} for the bond occupation we obtain
the corresponding Li--Sokal bounds \reff{Li_Sokal_bound_CH} \cite{Li_Sokal}. 

Similar bounds for the energy ${\cal E}$ are obtained by using the 
identity $E({\cal N} | \{ \sigma \} ) = p {\cal E}$ 
\cite{Li_Sokal,Salas_Sokal_Potts3}. This equation implies the following 
relation between the autocorrelation functions of the bond occupation and 
energy:
\be
\rho_{\cal EE}(t) = {\rho_{\cal NN}(t+1) \over \rho_{\cal NN}(1)} 
                  \geq \rho_{\cal NN}(t) \; .  
\ee
This implies in particular, the following set of equalities and inequalities 
\cite{Salas_Sokal_Potts3}:
\begin{subeqnarray}
\slabel{set_inequalities1}
\tau_{{\rm int},{\cal N}} &\geq& \tau_{{\rm int},{\cal E}} = 
 {\tau_{{\rm int},{\cal N}} - 1/2 \over \rho_{\cal NN}(1)} - {1\over 2} 
 \geq {\tau_{{\rm int},{\cal N}} \over \rho_{\cal NN}(1)} \; , \\
\tau_{\rm exp} &\geq& \tau_{{\rm exp},{\cal E}} = \tau_{{\rm exp},{\cal N}}\;. 
\slabel{set_inequalities2}
\label{set_inequalities}
\end{subeqnarray}
{}From \reff{set_inequalities} the Li--Sokal bounds \reff{Li_Sokal_bound_CH}
for the energy follow (and also for the exponential autocorrelation 
time $\tau_{\rm exp}$). We can also deduce from Eq.~\reff{set_inequalities1} 
that if $\rho_{\cal NN}(1) \neq 0$ as $L\rightarrow\infty$, then  
\cite{Salas_Sokal_Potts3}.
\be
z_{{\rm int},{\cal E}} =  z_{{\rm int},{\cal N}}  \; . 
\ee

\section{Swendsen--Wang--like algorithm for the Ashkin--Teller model} 
\label{sec_AT}

The AT model is a generalization of the Ising model 
to a four-state model. To each site $i$ of the lattice we assign  
two Ising spins $\sigma_i = \pm1$ and $\tau_i = \pm 1$; they interact 
through the Hamiltonian 
\be
{\cal H} = - J  \sum_{\<i,j\>} \sigma_i \sigma_j 
           - J' \sum_{\<i,j\>} \tau_i   \tau_j 
           - K  \sum_{\<i,j\>} \sigma_i \sigma_j \tau_i   \tau_j 
\label{hamiltonian_AT}
\ee
It can be regarded as two Ising models (with nearest-neighbor couplings $J$ and
$J'$) interacting via a four-spin coupling $K$. 
This model enjoys several symmetries: 1)  
It is symmetric under permutations of $(\sigma,\tau,\sigma\tau)$. This 
implies that \reff{hamiltonian_AT} is symmetric under permutations
of the couplings $(J,J',K)$. 
2) On any {\em bipartite} lattice, \reff{hamiltonian_AT} is symmetric 
under sublattice flip of two of the spins $(\sigma,\tau,\sigma\tau)$.
This means that \reff{hamiltonian_AT} is symmetric under sign flip of 
two of the couplings $(J,J',K)$. 

\begin{figure}[htb]
  \centering
  \begin{tabular}{c}
    \epsfxsize=200pt\epsffile{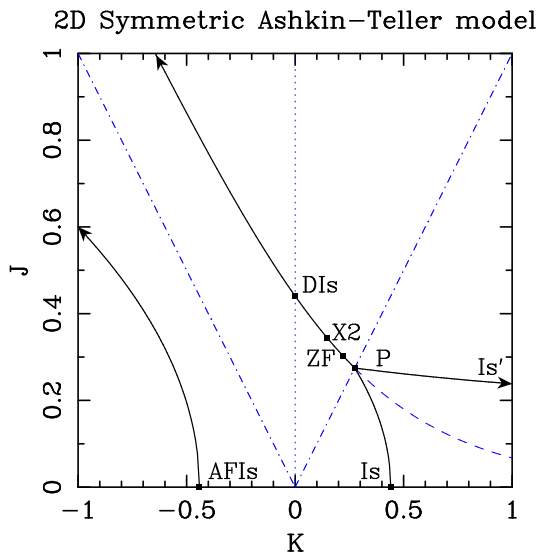}
  \end{tabular}
  \caption[a]{\protect\label{figure_AT_phase_diagram}
   The phase diagram of the square-lattice symmetric Ashkin--Teller model. 
   This model is symmetric under $J\rightarrow -J$, so we only depict the 
   half plane $J\geq 0$. The points DIs, X2, ZF, and P correspond to the 
   models analyzed in this paper (See text). 
   }
\end{figure}

In this paper we are mainly interested in one particular case of 
the Hamiltonian \reff{hamiltonian_AT}: the {\em symmetric} Ashkin--Teller 
(SAT) model  characterized by $J=J'$. 
Although we do not know how to solve this model analyticly, we have a 
fairly good understanding of its phase diagram on a square lattice 
\cite{Baxter} (See Figure~\ref{figure_AT_phase_diagram}). 
The line $K=0$ (dotted line in Figure~\ref{figure_AT_phase_diagram}) 
corresponds to two decoupled Ising models; its ferromagnetic critical point
is denoted by DIs. 
The line $J=K$ corresponds to the ferromagnetic 4-state Potts model, and its 
critical point is denoted by P ($J=K={1\over 4}\log 3$). 
The line $J=-K$ corresponds to the 4-state Potts antiferromagnet, which in 
known to be non-critical at all temperatures $T\geq 0$ \cite{Ferreira_Sokal}.  
The 4-state Potts subspace is depicted with dash-dotted lines in 
Figure~\ref{figure_AT_phase_diagram}.
The line $J=0$ corresponds to an Ising model in the variable $\sigma\tau$; 
its ferromagnetic critical point is denoted by Is, and its antiferromagnetic  
counterpart by AFIs. 
Finally, when $K=\infty$ we have another Ising model with $\sigma=\tau$, and 
coupling $2J$. Its ferromagnetic critical point is denoted by Is'. 

There are several critical curves in the SAT model. The most important one is 
the self-dual curve $e^{-2 K} = \sinh 2 J$. 
This curve is critical only for $K \leq {1\over 4}\log 3$ (solid curve in
Figure~\ref{figure_AT_phase_diagram}), and is noncritical for 
$K > {1\over 4}\log 3$ (dashed curve in Figure~\ref{figure_AT_phase_diagram}). 
The critical part belongs to the universality classes of the conformal field
theories with conformal charge $c=1$. Along this line the critical exponents 
vary continuously and they are known exactly 
\cite[and references therein]{Salas_Sokal_AT}. 
At the point P two more critical curves emerge: one goes to the critical 
point Is and the other one goes to the critical point Is' at $K=\infty$. 
Finally, there is another critical curve emerging from the critical point 
AFIs and pointing towards $K=-\infty$. The exact location of these three 
curves is unknown, although there are good numerical approximations  
\cite{Kamieniarz_97}. These lines are believed to belong to the 
Ising universality class.  

%
%

We have considered two different cluster algorithms for the AT model. In all 
cases we can assume that 
\be 
J, J' \geq |K| 
\label{constraints}
\ee
This condition is always achievable via the symmetries of the AT
Hamiltonian \reff{hamiltonian_AT} for a bipartite lattice. 

\medskip 

\noindent 
{\bf Direct algorithm:} The idea behind this algorithm is the same as in 
the standard SW algorithm for the Potts model: we decompose the Boltzmann 
weight by introducing new dynamical variables, and we then simulate the joint 
model by alternately updating one set of variables conditional on the other 
set. As we have two distinct sets of Ising spins, we expect to introduce
two distinct sets of auxiliary variables: $m_{ij},n_{ij}=0,1$ (associated
to the spins $\sigma$, $\tau$ respectively). 
The Boltzmann weight of the joint model is  
\begin{eqnarray}
 &  &
W_{\rm joint}(\{\sigma,\tau\};\{m,n\}) \;=\; {1 \over Z} \prod_{\<ij\>} \left[ 
    e^{-2(J+J')} \delta_{m_{ij},0} \delta_{n_{ij},0} \; + \nonumber \right. \\
& & \qquad \qquad
    e^{-2J'} \left[ e^{-2K} - e^{-2J} \right]
    \delta_{\sigma_i,\sigma_j}
    \delta_{m_{ij},1} \delta_{n_{ij},0} \;  + \nonumber \\
& & \qquad \qquad
    e^{-2J}  \left[ e^{-2K} - e^{-2J'} \right]
    \delta_{\tau_i,\tau_j} \delta_{m_{ij},0} \delta_{n_{ij},1} \; +
    \nonumber \\
& & \qquad \qquad \left. 
    \left[ 1 - e^{-2(J'+K)} - e^{-2(J+K)}
             + e^{-2(J+J')} \right]
    \delta_{\sigma_i,\sigma_j} \delta_{\tau_i,\tau_j}
    \delta_{m_{ij},1} \delta_{n_{ij},1} \right] 
    \;.\qquad
\label{Boltzmann_weight_joint}
\end{eqnarray}
Indeed, the Boltzmann weight of the original AT model is the marginal 
distribution of \reff{Boltzmann_weight_joint} with respect to the spin 
variables. From \reff{Boltzmann_weight_joint} we can read off the conditional 
probabilities needed in the SW procedure 
(See Ref.~\cite{Salas_Sokal_AT} for more details):

\medskip 

\noindent
{\bf Step 1.} Update the bond variables $\{m,n\}$ using the conditional 
              distribution $P_{\rm bond} = E(\cdot | \{\sigma,\tau\})$.  

\medskip 

\noindent
{\bf Step 2.} Update the spin variables $\{\sigma,\tau\}$ using the 
              conditional distribution $P_{\rm spin} = E(\cdot | \{m,n\})$. 

\bigskip 

This algorithm reduces on the ferromagnetic 4-state Potts subspace 
$J=K>0$ to the standard SW algorithm.  
Wiseman and Domany \cite{Wiseman_Domany} have introduced
essentially this same decomposition of the Boltzmann weight using a different  
derivation as ours. They then studied numerically the single-cluster version 
of this algorithm. Here we study the many-cluster (``SW'') version. 

\medskip

\noindent
{\bf Embedding algorithm:} The direct algorithm is perfectly legal, but
it is somewhat complicated to write the computer code for its Step 1 in an 
efficient way. This motivated the introduction of a variant algorithm in 
which we deal with only one kind of spin ($\sigma$ or $\tau$) at a time.
Consider the Boltzmann weight of a given bond $\<ij\>$, {\em conditional on the
$\{\tau\}$ configuration} (i.e., the $\tau$ spins are kept fixed): it is
\be
\label{Boltzmann_weight_sigma}
    W_{\rm bond}(\sigma_i,\sigma_j;\tau_i,\tau_j) \;=\;
                 e^{-2(J+K\tau_i\tau_j)} +
    \left[ 1 - e^{-2(J+K\tau_i\tau_j)} \right] \delta_{\sigma_i,\sigma_j} \; .
\ee
We can simulate this system of $\sigma$ spins using a standard SW algorithm.
The effective nearest-neighbor coupling
\be
J_{ij}^{\rm eff}  \;=\;  J + K \tau_i \tau_j
\label{effective_coupling}
\ee
is no longer translation-invariant, but this does not matter. The key point 
is that the effective coupling is always {\em ferromagnetic}, due to the
condition \reff{constraints}.
An exactly analogous argument applies to the $\{\tau\}$ spins when the
$\{\sigma\}$ spins are held fixed.
The embedding algorithm for the AT model has therefore two parts:

\medskip

\noindent
{\bf Step 1}: Given the $\{\tau\}$ configuration (which we hold fixed), 
  we perform a standard SW iteration on the $\sigma$ spins.
  The probability $p_{ij}$ arising in the SW algorithm takes the value
  $p_{ij} = 1 - \exp[-2(J+K\tau_i\tau_j)]$.

\medskip

\noindent
{\bf Step 2}: Given the $\{\sigma\}$ configuration (which we hold fixed), 
   we perform a standard SW iteration on the $\tau$ spins. In this case, 
   $p_{ij} = 1 - \exp[-2(J'+K\sigma_i\sigma_j)]$.

\bigskip

Wiseman and Domany \cite{Wiseman_Domany} also constructed an embedding
version of their single-cluster algorithm. Furthermore, they showed that,
in the single-cluster context, the direct and embedding algorithms
define the {\em same} dynamics\footnote{
  More precisely, this equivalence holds when the embedding algorithm is
  defined by making a {\em random} choice of Step 1 or Step 2 at each
  iteration.};
only the computer implementation is different. However, this equivalence 
does {\em not} hold for our many-cluster algorithm.
In the direct algorithm we have independent clusters of $\sigma$ spins and
$\tau$ spins that could be flipped simultaneously. In the embedding algorithm
we have at each step only one of the two types of clusters.

The embedding algorithm, due to its simplicity, is the one used in
our MC study of the square-lattice SAT model.
We expect that it lies in the same dynamic universality class as the
direct algorithm, on the grounds that one SW hit of $\{\sigma\}$ followed by
one SW hit of $\{\tau\}$ should be roughly equivalent to one joint hit of
$\{\sigma,\tau\}$. Of course, we do not expect the autocorrelation times for
the two algorithms to be {\em equal}, but we do expect them to be
asymptotically {\em proportional} as the critical point is approached.

The embedding algorithm does {\em not}\/ reduce to the standard SW
algorithm on the 4-state Potts subspace. However, we expect 
that they do belong to the same dynamic universality class. This fact has
been numerically verified by comparing the dynamic data for the 4-state Potts 
model with the embedding algorithm to those quoted in Ref.~\cite{Li_Sokal} 
(which correspond to the standard SW algorithm). We see that the
ratio of the two autocorrelation times is more or less constant within 
statistical errors, and that there is no systematic trend as $L$ grows. 
Thus, we conclude that they are proportional in the limit $L\rightarrow\infty$ 
\cite{Salas_Sokal_AT}: 
\be
{\tau_{{\rm int},{\cal E}}^{\rm direct} \over 
 \tau_{{\rm int},{\cal E}}^{\rm embedded} } = 1.516 \pm 0.035  
\ee
{}From Table~\ref{table_AT} we see that the performance of the embedding 
algorithm is reasonably good on the self-dual curve of the SAT model 
compared to the local algorithms.  

%
%
\begin{table}[htb]
\centering
\begin{tabular}{|l|l|l|}
\hline\hline
Model    & \multicolumn{1}{|c|}{$z_{{\rm int},{\cal E}}$} &
           \multicolumn{1}{|c|}{$\alpha/\nu$} \\
\hline\hline
X2      & $0.477 \pm 0.028$ \cite{Salas_Sokal_AT} & $\approx 0.4183$ \\
\hline
ZF      & $0.733 \pm 0.014$ \cite{Salas_Sokal_AT} & $2/3 \approx 0.666667$ \\
\hline\hline
\end{tabular}
\caption{\protect\label{table_AT}
Numerical estimates of the dynamical critical exponent
$z_{{\rm int},{\cal E}}$ for two points on the self-dual curve of the 
square-lattice symmetric Ashkin--Teller model. These points interpolate
between the Ising and 4-state Potts critical models (See 
Figure~\protect\ref{figure_AT_phase_diagram}). 
We also show the values of the ratio $\alpha/\nu$.
}
\end{table}

The direct algorithm for the AT model satisfies the Li--Sokal bound
\reff{Li_Sokal_bound_CH}. The proof is a straightforward generalization of the
one given in Section~\ref{sec_sw} for the Potts case 
\cite[Appendix A]{Salas_Sokal_AT}. As we expect that both the direct and the
embedding algorithms belong to the same universality class, we expect that
there is a Li--Sokal bound also for the embedding algorithm. In
Table~\ref{table_AT} we show the dynamic critical exponents and the values
of the ratio $\alpha/\nu$ for several points on the self-dual curve of the 
SAT model.
We have studied two points on the SAT self-dual curve between the critical 
Ising and 4-state Potts models; they are denoted by X2 and ZF 
(See Figure~\ref{figure_AT_phase_diagram}).  
X2 = $(J\approx 0.344132,K\approx 0.147920)$ is one of the points
considered in \cite{Wiseman_Domany}, and
ZF = $(J\approx 0.302923, K\approx 0.220343)$ corresponds to a model
studied by Zamolodchikov and Fateev \cite{Zamolodchikov}. In both cases
the Li--Sokal bound is satisfied and the difference
$z_{{\rm int},{\cal E}}-\alpha/\nu$ is not very large.

\section{Testing the sharpness of the Li--Sokal bound} \label{sec_results}

The most straightforward (and naive) method to study the sharpness of the 
Li--Sokal bound \reff{Li_Sokal_bound_CH}/\reff{Li_Sokal_bound}  
is to fit the autocorrelation time to a power-law 
$\tau_{\rm int} = A L^{z_{\rm int}}$ and compare the dynamic 
critical exponent $z_{\rm int}$ to the ratio $\alpha/\nu$. 
This procedure has a clear disadvantage: in some models the leading term 
of the specific heat is not a pure power law (e.g., in the 2D $q=2,4$ Potts 
models). And there is no reason why this should not happen for the 
autocorrelation times as well. This motivates the search for a different 
criterium: The Li--Sokal bound \reff{Li_Sokal_bound_CH} tells us that the  
ratio $\tau_{{\rm int},{\cal E}}/C_H \geq \hbox{const.}$, 
so we may try to fit it to different Ans\"atze.\footnote{We 
    have used a standard weighted least-squares
    method in all the fits presented here. 
    As a precaution against corrections to scaling,
    we impose a lower cutoff $L \ge L_{min}$
    on the data points admitted in the fit,
    and we study systematically the effects of varying $L_{min}$ on the
    parameter estimates and on the $\chi^2$ value.
    In general, our preferred fit corresponds to the smallest $L_{min}$
    for which the goodness of fit is reasonable
    (e.g., the confidence level is $\gtapprox$ 10--20\%),
    and for which subsequent increases in $L_{min}$ do not cause the
    $\chi^2$ to drop vastly more than one unit per degree of freedom.
}
In this section we will consider only the energy autocorrelation times, as 
this is one of the slowest modes of the system. We have considered 
three scenarios:

\begin{enumerate}   

\item The Li--Sokal bound is {\em sharp}: 
      $z_{{\rm int},{\cal E}} = \alpha/\nu$. 
      In this case we expect that 
      the ratio $\tau_{{\rm int},{\cal E}}/C_H$ will converge to a constant
      as $L\rightarrow\infty$. In particular, we have tried three different
      Ans\"atze compatible with this scenario 
\be 
{\tau_{{\rm int},{\cal E}} \over C_H} = \left\{ \begin{array}{l}
                 A \\
                 A + B L^{-\Delta} \\
                 A + B/ \log L 
                 \end{array} \right. 
\label{sharp_ansatz}
\ee
     The second (resp. third) Ansatz corresponds to power-law  
     (resp. additive logarithmic) corrections to scaling. 

\item The Li--Sokal bound is {\em sharp modulo a logarithm}: 
      $z_{{\rm int},{\cal E}} = \alpha/\nu \times \log^p$ with $p>0$. 
      In this case, the ratio $\tau_{{\rm int},{\cal E}}/C_H$ will 
      diverge like $\log^p L$. We have considered two Ans\"atze compatible 
      with this scenario
\be
{\tau_{{\rm int},{\cal E}} \over C_H} = \left\{ \begin{array}{l}
                 A + B \log L \\
                 A \log^p L
                 \end{array} \right.
\label{sharp_modulo_a_log_ansatz}
\ee
%

\item The Li--Sokal bound is {\em not sharp}:
      $z_{{\rm int},{\cal E}} = \alpha/\nu + p$ with $p> 0$. In this case the
      ratio $\tau_{{\rm int},{\cal E}}/C_H$ will diverge like a power law. 
      This scenario is represented by the Ansatz: 
\be
{\tau_{{\rm int},{\cal E}} \over C_H} = A L^{\alpha/\nu + p}
\label{non_sharp_ansatz}
\ee

\end{enumerate}

Our numerical results
\cite{Salas_Sokal_AT,Salas_Sokal_Potts3,Salas_Sokal_FSS,Salas_Sokal_Ising}
show that there are only two Ans\"atze that are likely
to the six models we have considered here 
(See Tables~\ref{table_results_ratio}--\ref{table_results_tau_int}).   
These are
\be
{\tau_{{\rm int},{\cal E}} \over C_H} = \left\{ \begin{array}{ll}
                 A + B \log L & \\
                 A L^p        & \qquad \mbox{\rm with $p$ small}
                 \end{array} \right.
\ee
Thus, the Li--Sokal bound seems to be either {\em sharp modulo a logarithm}
or {\em non-sharp} by a small power $0.05 \ltapprox p \ltapprox 0.12$.

%
%
\begin{table}[htb]
\centering
\begin{tabular}{lllrlcc}
\hline\hline
Model   & \multicolumn{1}{c}{Ansatz}
        & \multicolumn{1}{c}{Parameters}
        & \multicolumn{1}{c}{$L_{min}$}
        & \multicolumn{1}{c}{$\chi^2$}
        & $DF$ & level \\
\hline\hline
Ising   & $A + B/\log L$      & & 192      & $1.50$    & 1    & 22\% \\ 
        & $A\log L + B$       & &  96      & $1.47$    & 4    & 83\% \\ 
        & $A L^p$             & 
                 $p=0.0593(23)$ &  96      & $1.35$    & 4    & 85\% \\ 
\hline
$q=3$   & $A\log L + B$       & &  64      & $1.93$    & 3    & 59\% \\
        & $A L^p$             &
                 $p=0.084(2)$ &    32      & $1.72$    & 4    & 79\% \\
\hline
$q=4$   & $A\log L + B$       & &  16      & $1.99$    & 5    & 85\% \\
        & $A L^p$             &
                 $p=0.119(11)$&    16      & $1.06$    & 5    & 96\% \\
\hline
X2      & $A\log L + B$       & &  16      & $1.43$    & 4    & 84\% \\
        & $A L^p$             &
                 $p=0.051(9)$ &    16      & $1.28$    & 4    & 86\% \\
\hline
ZF      & $A\log L + B$       & &  16      & $1.03$    & 4    & 91\% \\
        & $A L^p$             &
                 $p=0.077(12)$&    16      & $1.23$    & 4    & 87\% \\
\hline\hline 
\end{tabular}
\caption{\protect\label{table_results_ratio}
We show the fits to the ratio $\tau_{{\rm int},{\cal E}}/C_H$ for the different 
models we consider in this paper. For each Ansatz, we give the value of 
$L_{min}$, the value of the $\chi^2$, the number of degrees of freedom ($DF$),
and the confidence level. For the power-law Ansatz $A L^p$ we also give 
the estimate of the power $p$. Note that for the 4-state Potts, X2 and ZF 
models, the error bar of the ratio $\tau_{{\rm int},{\cal E}}/C_H$ was
computed using the triangle inequality. Thus, the corresponding $\chi^2$ 
values are somewhat overestimated. 
}
\end{table}

%
%
\begin{table}[htb]
\centering
\begin{tabular}{llllrlcc}
\hline\hline
Model   & \multicolumn{1}{c}{Ansatz}  
        & \multicolumn{2}{c}{Parameters}  
        & \multicolumn{1}{c}{$L_{min}$} 
        & \multicolumn{1}{c}{$\chi^2$} 
        & $DF$ & level \\ 
\hline\hline
Ising   &  $A L^{z_{{\rm int},{\cal E}}}$ & 
           $z_{{\rm int},{\cal E}}$ & $=0.2265(50)$ & 
                                   192       & $0.0017$ & 1    & 97\%  \\
        & $A \log L + B$    & &  & 192       & $1.44$   & 1    & 23\%  \\
        & $A \log^2 L + B\log L $&    & 
                                 & 96       & $1.83$    & 4    & 77\%  \\ 
        & $A L^p \log L$         & 
          $p$ & $= 0.0504(23)$   & 96       & $1.35$    & 4    & 85\%  \\
\hline
$q=3$  & $A L^{z_{{\rm int},{\cal E}}}$ &
                    $z_{{\rm int},{\cal E}}$ & $=0.515(6)$ &
                                  128       & $0.44$    & 2    & 80\%  \\ 
       & $L^{2/5}(A \log L + B)$ & & &  
                                  32        & $1.54$    & 4    & 82\%  \\ 
\hline
$q=4$  & $A L^{z_{{\rm int},{\cal E}}}$ &
                    $z_{{\rm int},{\cal E}}$& $=0.876(11)$ &
                                  32        & $2.54$    & 4    & 64\%  \\
       & $L \log^{-3/2} L (A \log L + B)$ & &  &  
                                  16        & $1.53$    & 5    & 77\%  \\
       & $L^{1+p} \log^{-3/2} L$ & 
                 $p$ & $= 0.153(28)$ & 128     & $1.30$    & 2    & 52\%  \\ 
\hline 
X2     & $A L^{z_{{\rm int},{\cal E}}}$ &
                    $z_{{\rm int},{\cal E}}$ & $=0.477(28)$ &
                                  128       & $0.39$    & 1    & 53\%  \\
       & $L^{0.4183}(A\log L + B)$ & & &  
                                  128       & $0.42$    & 1    & 52\%  \\
\hline
ZF     & $A L^{z_{{\rm int},{\cal E}}}$ &
                    $z_{{\rm int},{\cal E}}$ & $=0.733(14)$ &
                                  32        & $1.48$    & 3    & 68\%  \\
       & $L^{2/3}(A\log L + B)$ & &  &
                                  16        & $1.53$    & 4    & 82\%  \\
\hline\hline 
\end{tabular}
\caption{\protect\label{table_results_tau_int}
We show the fits to the integrated autocorrelation times for the different 
models we consider in this paper. For each Ansatz, we give the value of 
$L_{min}$, the value of the $\chi^2$, the number of degrees of freedom ($DF$),
and the confidence level. For the power-law Ans\"atze we also give the
estimates of the powers ($p$ or $z_{{\rm int},{\cal E}}$). 
}
\end{table}

\begin{figure}[htb]
  \centering
  \begin{tabular}{cc}
    \epsfxsize=200pt\epsffile{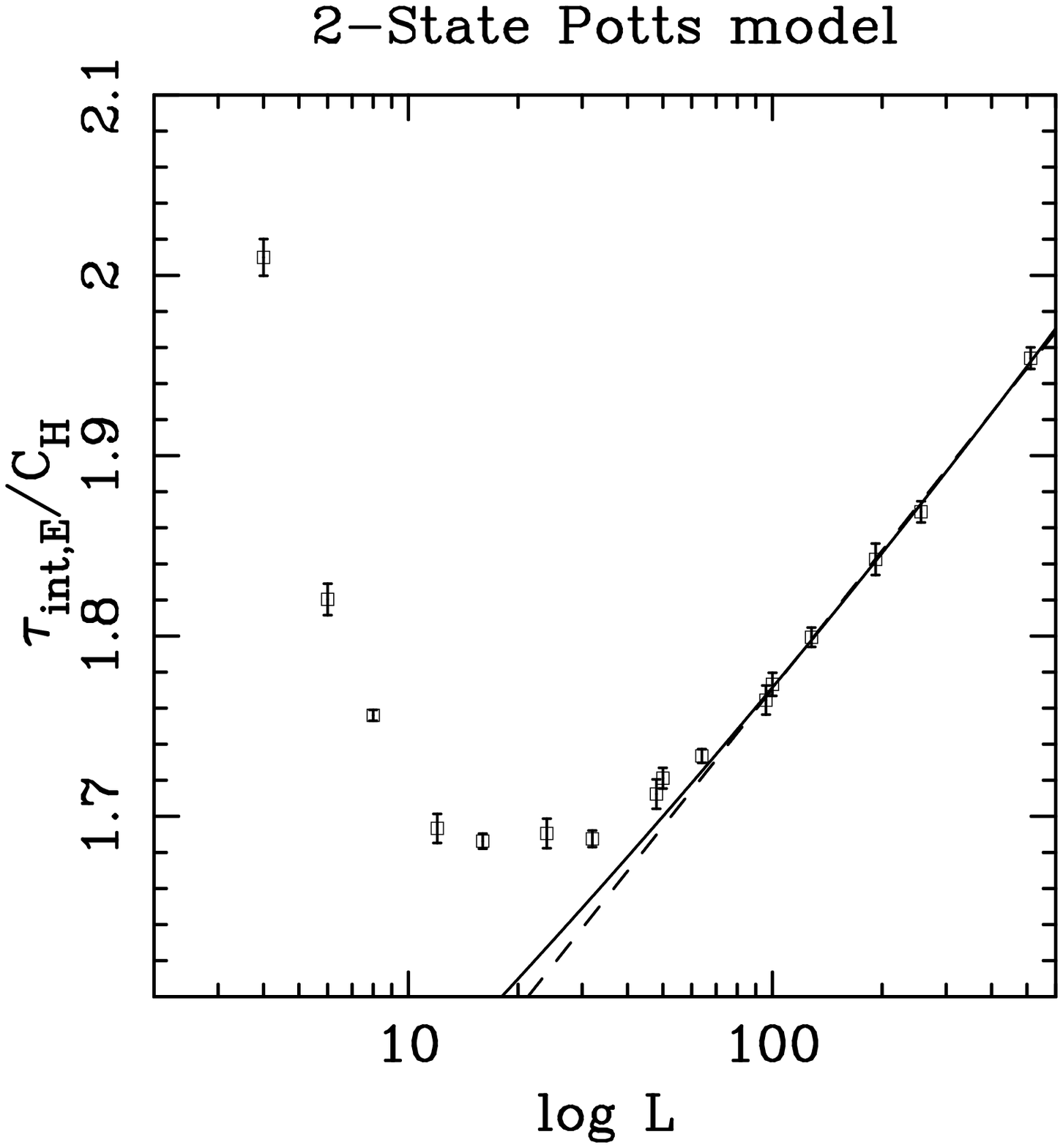} & 
    \epsfxsize=200pt\epsffile{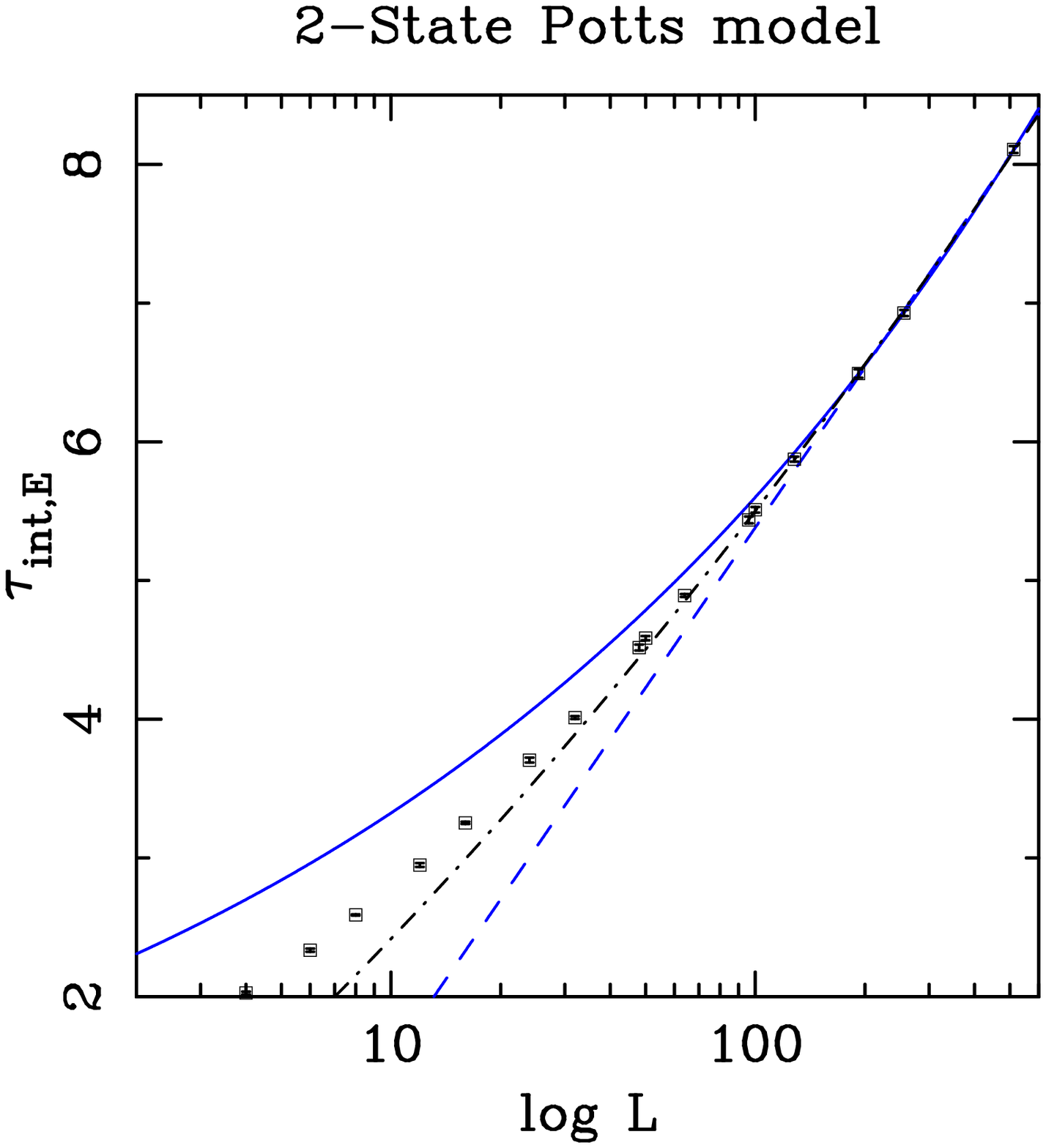} \\[2mm]
    (a)  & (b)   
  \end{tabular}
  \caption[a]{\protect\label{figure_ising}
    Dynamic critical behavior for the Ising model. 
    (a) Fits to the ratio $\tau_{{\rm int},{\cal E}}/C_H$. The solid line 
        shows the fit to the Ansatz $A L^p$; the dashed line shows the fit to  
        the Ansatz $A + B \log L$.  
    (b) Fits to the autocorrelation time $\tau_{{\rm int},{\cal E}}$.
        The dot-dashed curve shows the fits to the Ansatz 
        $A \log^2 L + B \log L$ (which is almost identical to the curve 
        coming from the Ansatz $A L^p \log L$). The dashed curve shows 
        the Ansatz $A+B\log L$;  
        and the solid curve, the Ansatz $A L^{z_{{\rm int},{\cal E}}}$. 
   }
\end{figure}

In the Ising model we have performed a high-precision MC simulation 
on lattices ranging from $L=4$ to $L=512$. In all cases we measured 
$10^6 \tau_{{\rm int},{\cal E}}$ iterations \cite{Salas_Sokal_Ising}.
In order to increase the statistics we merged our data to that of Baillie and
Coddington \cite{Baillie_91,Baillie_92}. 
{}From Table~\ref{table_results_ratio} we see that the sharp Ansatz
\reff{sharp_ansatz} is clearly disfavored: with a larger $L_{min}$ it
achieves a poorer confidence level. On the other hand, there is no
significant difference between the other two Ans\"atze
\reff{sharp_modulo_a_log_ansatz}/\reff{non_sharp_ansatz}. This can be seen
clearly in Figure~\ref{figure_ising}(a).
If we look at Table~\ref{table_results_tau_int}, we see that the pure-power-law
Ansatz is disfavored, as it is the logarithmic Ansatz proposed by several
authors \cite{Heermann_90,Baillie_91,Baillie_92}. Both of them need a larger
value of $L_{min}$ than in the other two Ans\"atze. Finally, the
non-sharp-by-a-power Ansatz \reff{non_sharp_ansatz} is now slightly favored
over the non-sharp-by-a-logarithm Ansatz \reff{sharp_modulo_a_log_ansatz},
but the difference is again probably not significant.
In Figure~\ref{figure_ising}(b) we see clearly that both fits are clearly
favored over the simple power-law and the logarithmic growth.

%
%

In the 3-state Potts model we have performed a MC simulation on lattices 
ranging from $L=4$ to $L=1024$. In all cases we measured 
$10^6 \tau_{{\rm int},{\cal E}}$ iterations \cite{Salas_Sokal_Potts3}.
{}From Table~\ref{table_results_ratio} we see that the power-law Ansatz  
\reff{non_sharp_ansatz} to the ratio $\tau_{{\rm int},{\cal E}}/C_H$ is  
slightly favored over the logarithmic Ansatz \reff{sharp_modulo_a_log_ansatz}. 
However, if we look at Table~\ref{table_results_tau_int} we observe that 
the Ansatz $L^{2/5}(A \log L + B)$ is favored over the 
non-sharp-by-a-power Ansatz: the former achieves a similar value of the 
confidence level with a smaller value of $L_{min}$. 
In this model the finite-size corrections in the specific heat are very 
important \cite{Salas_Sokal_Potts3} and this is possibly the cause that one 
scenario is favored when we study $\tau_{{\rm int},{\cal E}}$, and a  
different one is favored when we consider the ratio 
$\tau_{{\rm int},{\cal E}}/C_H$.

In the 4-state Potts model we performed a MC simulation using the  
embedding algorithm for the SAT model
on lattices ranging from $L=16$ to $L=1024$. In all cases (but $L=1024$), we
measured at least $10^4 \tau_{{\rm int},{\cal E}}$ iterations; for $L=1024$ we
could only measure $1500 \tau_{{\rm int},{\cal E}}$ iterations 
\cite{Salas_Sokal_FSS}.
If we look at Table~\ref{table_results_ratio} we see that the power-law fit
is favored over the logarithmic fit: for the same value of $L_{min}$ the 
latter gets twice as much $\chi^2$ than the former. However, if we look at 
the fits of $ \tau_{{\rm int},{\cal E}}$ (Table~\ref{table_results_tau_int}),  
we conclude that the non-sharp-by-a-logarithm Ansatz is favored over the 
other two (i.e., the non-sharp-by-a-power and the pure power-law scenarios). 
With a smaller value of $L_{min}$ it achieves the best confidence level of all
three.  

Finally, the models X2 and ZF were studied using the embedding algorithm for 
SAT model. The simulations were performed on lattices ranging from $L=16$ to 
$L=512$. We performed at least $10^4 \tau_{{\rm int},{\cal E}}$ measurements
for the ZF model, and $3 \times 10^4 \tau_{{\rm int},{\cal E}}$ for the X2 
model. In Tables~\ref{table_results_ratio}-\ref{table_results_tau_int} we 
see almost no difference between the fits corresponding to the 
non-sharp-by-a-power \ref{non_sharp_ansatz} and the 
non-sharp-by-a-logarithm \ref{sharp_modulo_a_log_ansatz} Ans\"atze.  

In all cases we conclude that the Li--Sokal bound
\reff{Li_Sokal_bound_CH}/\reff{Li_Sokal_bound} is {\em not sharp} by a small
quantity: either a logarithm or a small power $p$. Furthermore, there is 
some kind of continuity among the values of the power $p$: it grows from 
$q \approx 0.05$ in the Ising model to $q\approx 0.12$ in the 4-state model,
irrespective if we go through the 3-state Potts model ($p\approx 0.08$) 
or through the SAT self-dual curve ($p(\hbox{\rm X2}) \approx 0.05$ and 
$p(\hbox{\rm ZF}) \approx 0.08$). 
It is extremely difficult to distinguish between these two scenarios with 
lattices up to $L=1024$.
We need larger lattices $L\gg1024$ in order to solve this question. 

It is intriguing why the Li--Sokal bound is so close to be sharp in two 
dimensions, and clearly non-sharp for $d>2$. The explanation is unknown; but 
a reasonable guess is that in $d>2$ the energy and the bond occupation are not 
among the slowest modes of the system (as in $d=2$). It would be very 
interesting to characterize such slow modes for $d>2$.

\section{Exponential autocorrelation times} \label{sec_tau_exp}

So far we have analyzed the behavior of the integrated autocorrelation time.
However, it is not guaranteed that this behavior coincides with that of the 
exponential autocorrelation time. In general, we expect that 
$z_{{\rm int},A} \neq z_{{\rm int},A}$ for most observables $A$  
\cite{Sokal_Cargese}. This issue can be explained using the scaling relation
for the autocorrelation function \reff{autocorrelation_function} 
\be
\rho_{AA}(t;L) \approx |t|^{-p_A} h_A \left[ {t \over \tau_{{\rm exp},A}}; 
               {\xi(L) \over L} \right] 
\label{scaling_relation_rho}  
\ee
where $p_A$ is an exponent, $h_A$ is a scaling function, and $\xi$ is the 
correlation length of the system. Summing over the time $t$, we obtain 
\begin{subeqnarray}
\tau_{{\rm int},A} &\sim&  \tau_{{\rm exp},A}^{1 - p_A} \\
z_{{\rm int},A}    &=&     (1 - p_A) z_{{\rm exp},A} 
\end{subeqnarray}
Thus, both dynamic critical exponents are different unless $p_A=0$. If this is
the case, then the scaling relation \reff{scaling_relation_rho} simplifies
to
\be
\rho_{AA}(t;L) \approx \widehat{h}_A \left[ {t \over \tau_{{\rm int},A}};
               {\xi(L) \over L} \right] 
\label{scaling_relation_rho_pA=0}
\ee
where $\widehat{h}_A$ is another scaling function. This Ansatz can be 
numerically verified by plotting $\rho_{\cal EE}$ as a function of 
$t/\tau_{{\rm int},{\cal E}}$ and testing if the data coming from 
different values of $L$ collapse well onto a single curve. If so, then 
$p_{\cal E}=0$. 

\begin{figure}[htb]
  \centering
   \begin{tabular}{c}
     \epsfxsize=200pt\epsffile{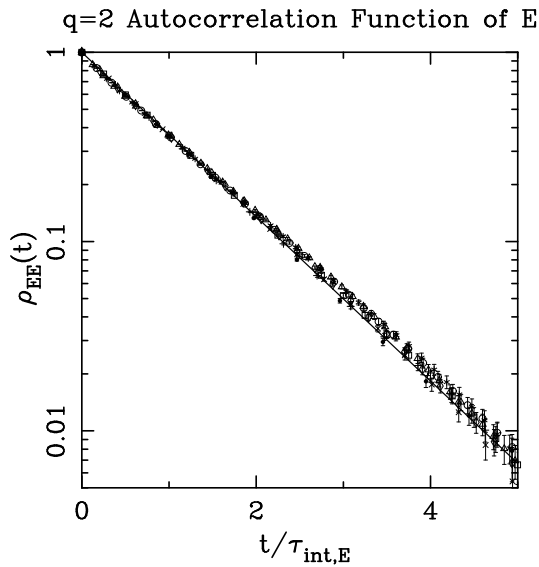} 
   \end{tabular}
   \caption[a]{\protect\label{figure_ising_tau_exp}
      Plot of $\rho_{\cal EE}(t)$ versus $t/\tau_{{\rm int},{\cal E}}$ for the
      Ising model. The different symbols denote the different lattice sizes:
      $L=4$ ($\bullet$), $L=8$ ($+$), $L=16$ ($\times$), 
      $L=32$ ($\Box$), $L=64$ ($\diamondsuit$), $L=128$ ($\circ$), 
      $L=256$ ($\ast$), and $L=512$ ($\triangle$). We have also depicted the
      line corresponding to a pure exponential 
      $\rho_{\cal EE}(t)=\exp(-t/\tau_{{\rm int},{\cal E}})$. 
   }
\end{figure}

We have plotted $\rho_{\cal EE}$ versus $t/\tau_{{\rm int},{\cal E}}$ for 
the Ising model in Figure~\ref{figure_ising_tau_exp}. Similar plots 
for the 3- and 4-state Potts models can be found in 
Refs.~\cite{Salas_Sokal_Potts3,Salas_Sokal_AT}. In all cases we 
find that $\rho_{\cal EE}$ is close to a pure exponential and  
that the Ansatz \reff{scaling_relation_rho_pA=0} is satisfied 
within statistical errors (usually we find small corrections to scaling 
for the smaller values of $L$; e.g., in the Ising case the Ansatz 
\reff{scaling_relation_rho_pA=0} is satisfied for $L\geq 64$). Thus, 
the conclusion is that the exponential and integrated autocorrelation
times for the energy are likely to have the same critical exponent
\be
z_{{\rm int},{\cal E}} = z_{{\rm exp},{\cal E}}
\ee
for the SW algorithm (Potts model) and for the embedding algorithm (SAT model). 

The value of $\tau_{{\rm exp},{\cal E}}$ can in principle be obtained 
by fitting $\rho_{\cal EE}$ to an exponential. However, this is not an 
easy task as the MC estimates of $\rho_{\cal EE}$ for different $t$ are highly 
correlated \cite{Salas_Sokal_AT}. One should compute the full covariance 
matrix for these random variables. In Ref.~\cite{Salas_Sokal_AT} a method 
was proposed to perform this computation. However, high-precision data is
needed in order to obtain meaningful results: we need at least 
$6\times 10^4 \tau_{{\rm int},{\cal E}}$ measurements. 
For all the models considered here we can nevertheless obtain crude estimates 
of the ratio $\tau_{{\rm int},{\cal E}}/\tau_{{\rm exp},{\cal E}}$, which 
are slightly smaller than 1.

%
%

%
%
\section*{Acknowledgments}

We wish to thank Alan Sokal for many useful discussions 
and helpful advice over the years. We would like also to thank Paul Coddington
for communicating to us his unpublished data; Andea Pelissetto for 
collaborating in the derivation of the direct algorithm for the 
Ashkin--Teller model and Doug Toussaint for helpful correspondence.  
The authors' research was supported in part by CICyT grant AEN97-1680.

%
%
%
%
%

\end{document}